\documentclass[a4paper,fleqn]{cas-dc}
\usepackage[numbers]{natbib}

\usepackage{arydshln}
\def\tsc#1{\csdef{#1}{\textsc{\lowercase{#1}}\xspace}}
\tsc{WGM}
\tsc{QE}
\tsc{EP}
\tsc{PMS}
\tsc{BEC}
\tsc{DE}

\begin{document}
\let\WriteBookmarks\relax
\def\floatpagepagefraction{1}
\def\textpagefraction{.001}

\shorttitle{Learnable Weight Initialization for  Volumetric Medical Image Segmentation}

\shortauthors{Shahina Kunhimon et~al.}

\title [mode = title]{Learnable Weight Initialization for Volumetric Medical Image Segmentation}                      



%
\author[1]{Shahina Kunhimon}[
                        orcid=0009-0001-6809-2285]

\cormark[1]

\fnmark[1]

\ead{shahina.kunhimon@mbzuai.ac.ae}



\affiliation[1]{organization={Mohammed Bin Zayed University of Artificial Intelligence },
    city={Abu Dhabi},
    country={UAE}}
\author[1]{Abdelrahman Shaker}
\author[1]{ Muzammal Naseer}



\author[1]{Salman Khan}
\author[1,2]{Fahad Shahbaz Khan}
\affiliation[2]{organization={Linkoping University},
    country={Sweden}}

\cortext[cor1]{Corresponding author}
\fntext[fn1]{This is the first author footnote.}

\begin{abstract}
Hybrid volumetric medical image segmentation models, combining the advantages of local convolution and global attention, have recently received considerable attention. While mainly focusing on architectural modifications, most existing hybrid approaches still use conventional data-independent weight initialization schemes which restrict their performance due to ignoring the inherent volumetric nature of the medical data. To address this issue, we propose a learnable weight initialization approach that utilizes the available medical training data to effectively learn the contextual and structural cues via the proposed self-supervised objectives. Our approach is easy to integrate into any hybrid model and requires no external training data. Experiments on multi-organ and lung cancer segmentation tasks demonstrate the effectiveness of our approach, leading to state-of-the-art segmentation performance. Our proposed data-dependent initialization approach performs favorably as compared to the Swin-UNETR model pretrained using large-scale datasets on multi-organ segmentation task. Our source code and models are available at: \url{https://github.com/ShahinaKK/LWI-VMS}.

\end{abstract}

\begin{keywords}
Hybrid architecture \sep Volumetric medical segmentation \sep Weight initialization 
\end{keywords}
\maketitle
\section{Introduction}
\label{sec:introduction}
In medical image segmentation, target organs and tissues are pixel-wise classified enabling better diagnosis, and treatment planning. Advances in deep learning methods have significantly improved medical image segmentation tasks, such as tumor \cite{swinunet}, \cite{SWINUNETR} and skin lesion \cite{modelgenesis} segmentation. Various successful convolutional neural network (CNN) models, self-attention (SA) based transformer models, and their combinations have been adapted for medical image segmentation tasks. Generally, it is necessary to have a large amount of annotated training data to achieve promising results with deep neural networks \cite{ViTs,thawkar2023xraygpt}. However, it is a complex and expensive process to collect and annotate medical images to curate large-scale benchmark datasets. The ethical and legal constraints associated with medical data to preserve the privacy and security of sensitive patient information make the data collection and annotation tasks more challenging. Therefore, the majority of the existing medical image segmentation methods focus on improving the architecture of deep neural networks.

The recent developments in vision transformers (ViTs) \cite{ViTs}, \cite{khan2022transformers}, \cite{vaswani2017attention} have enabled a hybrid design  \cite{SWINUNETR}, \cite{UNETR} incorporating the complementary properties of convolutional networks and self-attention based vision transformers for volumetric medical segmentation. However, we observe that these hybrid CNN-transformer models are typically initialized using conventional \textit{data-independent} weight initialization schemes \cite{glorot2010}, \cite{he2015} which can affect their overall segmentation performance. For example, the model training can converge to different solutions based on the weight initialization scheme employed as discussed in Section \ref{data_ind}.

In this work, we argue that self-supervised inductive biases that can capture the nature of volumetric data are likely to perform better than the conventional weight initialization schemes that are data-independent. To this end, we introduce a \textit{learnable weight initialization} approach that strives to explicitly exploit the volumetric nature of the medical data to induce contextual cues within the model at an early stage of training. These contextual cues are learned using our proposed self-supervised objectives. The segmentation models are based on encoder-decoder network design. Therefore, to learn contextual cues from a given volumetric input, our approach encourages the encoder to predict the correct order of shuffled sub-volumes while training the decoder to reconstruct the masked organs or part of an organ (Section.\ref{data_dep}).
As a result, data-dependent priors about the input structure can be effectively captured within the model weights across different scans of the volumetric input, resulting in better segmentation performance. Our contributions can be summarized as follows:

 \begin{itemize}
     \item We propose a learnable weight initialization method that can be integrated into any hybrid volumetric medical segmentation model to effectively train small-scale datasets.
     \item To learn such a weight initialization, we propose data-dependent self-supervised objectives tailored to learn the structural and contextual cues from the volumetric medical image datasets.
     \item We demonstrate the effectiveness of our approach by conducting experiments for multi-organ and tumor segmentation tasks, achieving superior segmentation performance without requiring additional external training data.
    \item  Our proposed weight initialization scheme, which relies solely on the training dataset at hand
    yields favorable results when compared to the single model performance of Swin-UNETR large-scale self-supervised pretraining \cite{tang2022self} on multi-organ segmentation task.
 \end{itemize}

 \begin{figure*}[t]
\setlength{\belowcaptionskip}{-2pt}
\setlength{\abovecaptionskip}{-2pt}
\setlength{\abovedisplayskip}{-2pt}
\setlength{\belowdisplayskip}{-2pt}
\begin{center}
\includegraphics[width=0.96\linewidth]{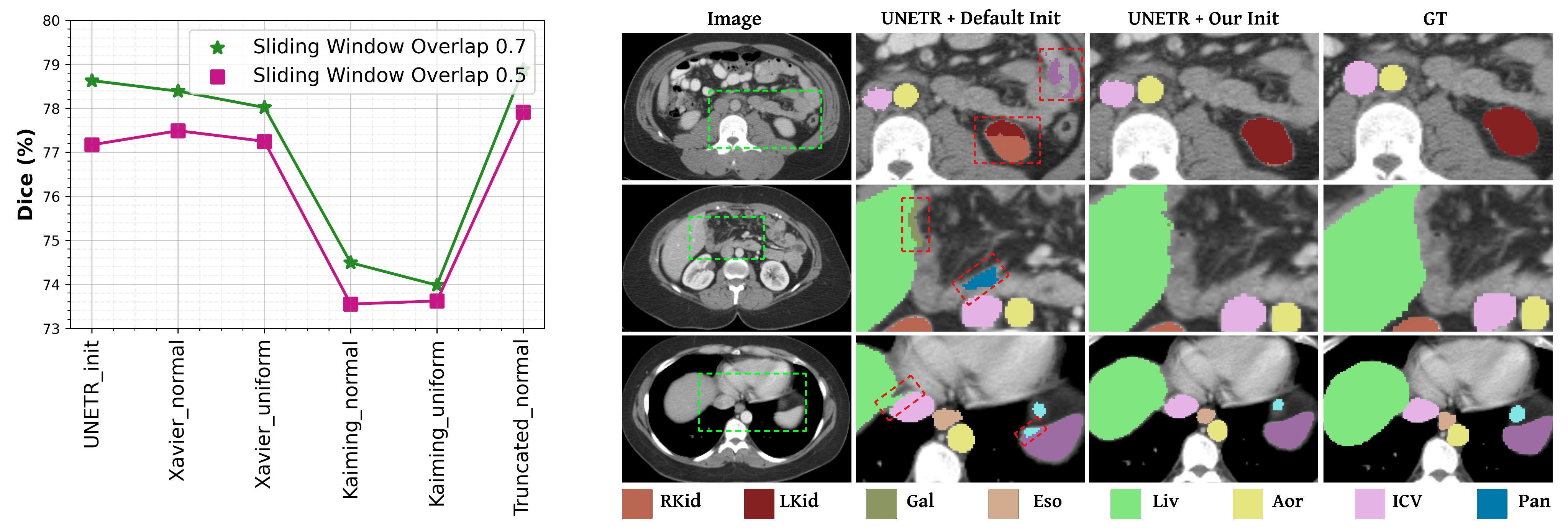}
\end{center}
\caption{\textbf{Left:} UNETR \cite{UNETR} is sensitive to different data-independent weight initialization schemes. We observe that UNETR performance drops significantly when initialized with the Kaiming normal method. Further, the truncated normal method gives better results than the default UNETR initialization. \textbf{Right:} Qualitative comparison on Synapse dataset results between the default and our proposed initialization (Init) method within the same UNETR framework. We enlarge the segmented area (green dashed boxes in column 1). Our method reduces the \textit{false positives} for organs compared to standard UNETR (red dashed box in column 2). Organs are shown in the legend below the examples. Best Viewed zoomed in.}
\label{fig:Diff_init}
\end{figure*}

\section{Related Work}
Medical image segmentation using deep learning techniques has garnered significant interest in healthcare research. These techniques can be broadly categorized into three groups: CNN-based, transformer-based, and hybrid approaches. 

A variety of models incorporating encoder-decoder structures with diverse CNN backbones have been adopted for medical image segmentation tasks. Deeplab \cite{Deeplab}, Fully Convolutional Networks (FCN) \cite{FCN}, and U-Net \cite{UNet} were some of them. Since the introduction of the U-Net \cite{UNet}, various CNN-based approaches \cite{cai2020dense}, \cite{3dunet}, \cite{huang2020unet}, \cite{nnUnet}, \cite{milletari2016v} have been introduced to extend the typical U-Net architecture for different medical image segmentation tasks. However, these CNN-based models cannot capture long-range correlations in the data due to the intrinsic locality of convolution operations which limits their performance in challenging segmentation problems.

Due to the success of the vision transformer models (ViTs), recent works have focused on investigating their applicability to medical segmentation tasks \cite{lahoud20223d}. For the volumetric medical image segmentation task, pure transformer-based designs were explored in \cite{swinunet} and \cite{karimi2021convolution}. Despite having the capability to capture the global structure via self-attention, ViTs require pre-training on large-scale datasets to inherent inductive biases and achieve promising performance \cite{awais2023foundational}, thereby limiting their adoption in medical imaging datasets because of the scarcity of the data. 

Several recent methods \cite{SWINUNETR}, \cite{UNETR}, \cite{UNETR++}, \cite{TransBTS}, \cite{nnFormer} have explored hybrid architectures with convolutional layers to encode CNN inductive biases and the self-attention layers for better global representation. UNETR \cite{UNETR} is a hybrid method for 3D medical segmentation tasks and is composed of a “U-shaped” encoder-decoder architecture, with a ViT transformer encoder to encode enriched global representation and a convolutional decoder. Swin UNETR \cite{SWINUNETR} adapted hierarchical Swin transformer as the vision encoder backbone to mitigate the drawbacks of fixed token size in the ViT encoder in UNETR architecture. nnFormer \cite{nnFormer} follows a hierarchical encoder-decoder architecture with a combination of interleaved convolution and self-attention operations which make use of both local and global volume-based self-attention mechanisms to encode the volume representations. UNETR++ \cite{UNETR++} extended the UNETR architecture by replacing the fixed transformer representation with a hierarchical efficient paired attention module to reduce the model complexity significantly.

Inspired by the success of the ConvNeXt architecture \cite{liu2022convnet} in various computer vision tasks, which integrates the ability of transformers to learn long-range dependencies into convolutional networks, several ConVNeXt-based volumetric medical segmentation networks, such as 3D-UX-Net \cite{lee20223d} and MedNeXt \cite{roy2023mednext} were introduced. In 3D-UX-Net \cite{lee20223d}, the Swin Transformer block from \cite{SWINUNETR} was replaced with ConvNeXt blocks, whereas MedNeXt \cite{roy2023mednext} follows a fully ConVNeXt based encoder-decoder architecture designed for volumetric medical image segmentation. Also, it offers four different configurations: MedNeXt-S, MedNeXt-B, MedNeXt-M, and MedNeXt-L, each with 2 kernel sizes (k=3 and k=5).

\begin{figure*}[!t]
\setlength{\belowcaptionskip}{-2pt}
\setlength{\abovecaptionskip}{-2pt}
\setlength{\abovedisplayskip}{-2pt}
\setlength{\belowdisplayskip}{-2pt}
\begin{center}
\includegraphics[width=0.96\linewidth]{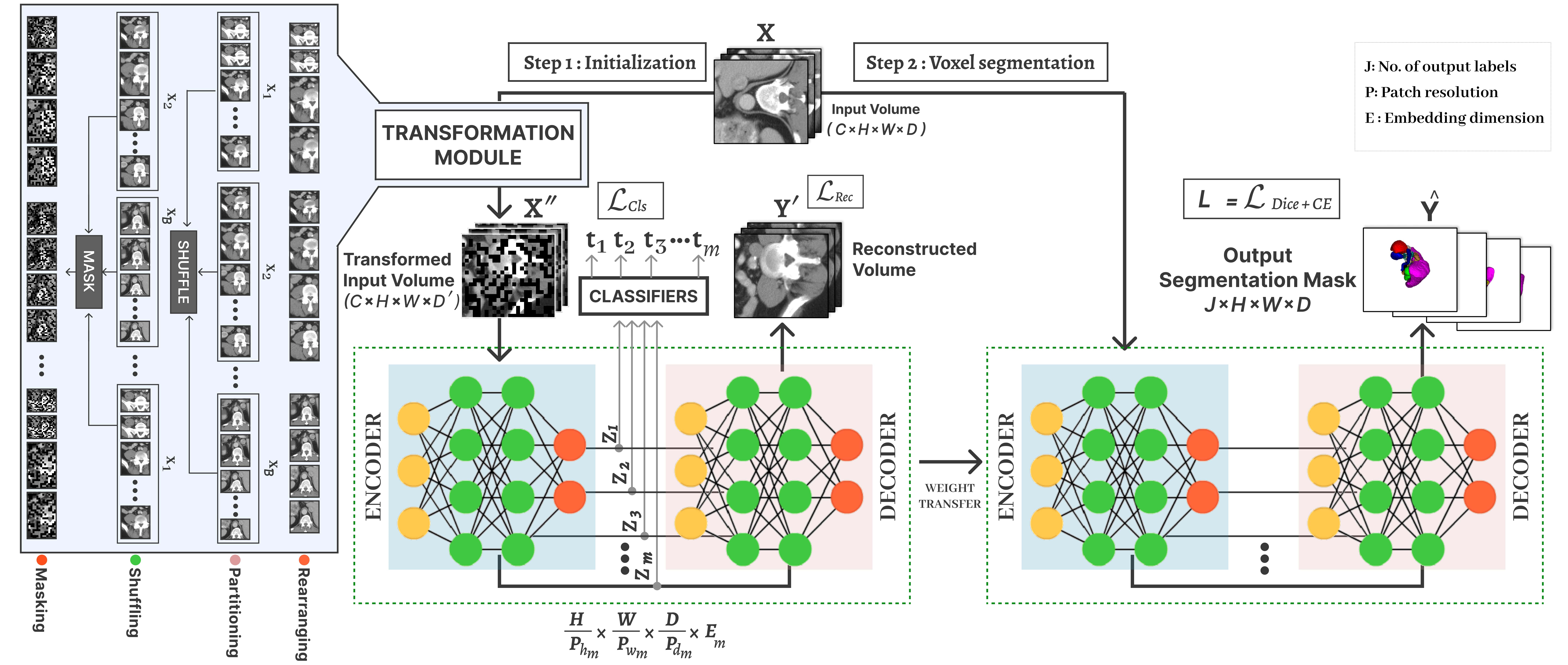}
\end{center}
   \caption{\textbf{Overview of our proposed approach:} To learn weight initialization using self-supervised tasks defined by the volumetric nature of the medical data. In the early stage of training (Step-1), we define the order prediction task within the encoder latent space, while simultaneously the decoder has to reconstruct the missing organs from masked $\&$ shuffled input. The masked $\&$ shuffled input is the result of our transformation module with 4 stages
 : depth-wise rearranging, partitioning into equal size sub-volumes, random shuffling of sub-volumes for the order prediction objective, and finally masking shuffled volume for the reconstruction objective. This allows the model 
 to learn structural and contextual consistency about the data 
 that provides an effective initialization for the segmentation task (Step-2). Our approach does not rely on any extra data 
 and therefore remains as computationally effective as the baseline while enhancing the segmentation performance.}
\label{fig:main_demo}
\end{figure*}

\subsection{Weight Initialization schemes}

Weight initialization plays a crucial role in deep neural network training, as it can have a strong impact on the training time as well as the quality of the resulting model. The objective of an initializer is to determine the initial network parameter values within a suitable region of the optimization landscape so that training converges to optimal solution \cite{li2018visualizing}. Random initialization is the most commonly used approach where the initial weights are assigned by randomly sampling from a given distribution such as standard normal and uniform distributions. Another method called truncated normal initializes weights through sampling from a normal distribution, similar to standard normal initialization. However, if the values fall outside a given range, they are truncated and resampled to be within the limits. Compared to the standard normal initialization method, this approach provides improved control over the initialization range, which is beneficial when considering prior knowledge or domain-specific constraints regarding the acceptable range of parameter values.

\begin{table*}[!t]\normalsize
    \centering
    \caption{\textbf{Baseline Comparison on Synapse dataset:} Our approach significantly improves UNETR on Synapse. Specifically, in terms of dice score, we observe significant improvements in small organs such as the aorta, gallbladder, and pancreas. The p-value is derived by comparing the average dice scores obtained over five runs of our proposed method and its corresponding baseline experiments. FPR and TNR correspond to the average False Positive Rate and True Negative Rate, respectively.}
    \label{table:UNETR synapse}
    \setlength{\tabcolsep}{5.5pt}
        \begin{tabular}{l | c c c c c c c c c | c | c}
            \hline
            \multicolumn{1}{c|}{\textbf{Method}}  & \multicolumn{9}{c|}{\textbf{Dice score (DSC)} $\uparrow$} & \textbf{FPR}  & \textbf{TNR}  \\
            \cline{2-10}
            & Spl & Rkid & Lkid & Gall & Liv & Sto & Aor &  Pan & Average & $\downarrow$ & $\uparrow$ \\
            \hline
            UNETR~\cite{UNETR} & \textbf{88.58} & 80.03 & 78.87 & 62.51 & 95.45 & 74.44 & 84.79 & 52.70 & 77.17 & 3.89e-05 & 0.99952\\
            \textbf{UNETR (Ours)} & 86.72 & \textbf{82.86} & \textbf{85.41} & \textbf{65.15} & \textbf{95.56} & \textbf{75.23} & \textbf{88.07} & \textbf{58.85} & \textbf{79.73} & {3.23e-05} & {0.99975}\\
            \cline{2-10}
             & \multicolumn{9}{|c|}{p-value  = 5.36e-04 < 0.01} & &\\
            \hline
        \end{tabular}
\end{table*}

Xavier initialization introduced in \cite{glorot2010}, also known as Glorot initialization, initializes the weights by sampling from a uniform or normal distribution with its standard deviation dependent on the number of input and output connections. This technique focuses on keeping the variance of the activations and gradients relatively constant during forward and backward propagation. In the Xavier uniform method, the range of the values for weight initialization is calculated using a uniform distribution $\mathcal{U}(-a, a)$, where the range limit $a$ is given by:
 \begin{equation}
\label{eq: xavier_uniform}
    a = G* \sqrt{\frac{2}{c_{in} +c_{out}}}
\end{equation}
For the Xavier initialization method using normal distribution $\mathcal{N}(0,\sigma^2)$, the standard deviation $\sigma$ is given by:
 \begin{equation}
\label{eq: xavier_normal}
    \sigma = G * \sqrt{\frac{2}{c_{in} +c_{out}}}
\end{equation}
In equations \ref{eq: xavier_uniform} and \ref{eq: xavier_normal}, $G$ corresponds to an optional scaling factor and $c_{in}$ and $c_{out}$ represents the number of previous layer (input) and current layer (output) connections respectively. 
Kaiming He Initialization \cite{he2015} is a variant of Xavier initialization introduced to mitigate the issue of vanishing gradients associated with the nonlinear activations by adjusting the distribution based on the number of inputs to the current layer. In the Kaiming uniform method, the values for weight initialization are based on a uniform distribution $\mathcal{U}(-b, b)$ bounded by the limit $b$ which is given by:
 \begin{equation}
\label{eq: kaiming_uniform}
    b = G * \sqrt{\frac{3}{c_{in}}}
\end{equation}
For the Kaiming initialization method using normal distribution $\mathcal{N}(0,\sigma^2)$, the standard deviation $\sigma$ is given by:

\begin{equation}
\label{eq: kaiming_normal}
    \sigma = \frac{G}{\sqrt{c_{in}}}
\end{equation}

 $G$ corresponds to an optional scaling factor and $c_{in}$ represents the number of input connections.

Generally, standard data-independent weight initialization techniques are adopted for medical imaging tasks. However, medical image datasets are very different from natural image datasets with respect to the variabilities in terms of imaging modalities and anatomical structure. Also, the region of interest (tumors or any structural abnormality) is relatively rare compared to the background or normal regions in the 3D medical image scans. Hence, employing specific data-dependent weight initialization schemes tailored for medical image segmentation tasks can assist the model in learning more meaningful representations by incorporating prior knowledge about the variability in the imaging modalities and object anatomy. This approach reduces the bias towards dominant classes, ultimately enhancing the segmentation outcome.

Pretraining on large-scale datasets is a popular data-dependent initialization approach explored across various application fields of deep learning. For volumetric medical image segmentation, pretraining on large-scale natural image datasets cannot guarantee good generalization due to the difference in the image distribution. Large-scale pretraining on medical datasets is not favorable since annotated medical data is deficient.

\begin{table*}[!t]\normalsize
\centering
\caption{\textbf {SOTA comparison on Synapse dataset:} We observe a large variance in the performance of existing methods across different organs. In comparison, our approach consistently performs better while increasing the overall performance. The p-values are computed using the average dice scores from five runs of our approach and its corresponding baseline. FPR and TNR correspond to the average False Positive Rate and True Negative Rate, respectively.}
\label{table: SOTA_synapse}
\setlength{\tabcolsep}{5.5pt}
\scalebox{0.9}{
    \begin{tabular}{l| c c c c c c c c c | c | c}
    \hline
    \multicolumn{1}{c|}{\textbf{Method}}  & \multicolumn{9}{c|}{\textbf{Dice score (DSC)} $\uparrow$} & \textbf{FPR}  & \textbf{TNR}\\
    \cline{2-10}
    & Spl & Rkid & Lkid & Gall& Liv & Sto & Aor&  Pan & Average &$\downarrow$ & $\uparrow$\\
    \hline
    U-Net \cite{UNet} & 86.67 & 68.60 & 77.77 & 69.72 & 93.43 & 75.58 & 89.07 & 53.98 & 76.85  &{-} &{-}\\
    TransUNet \cite{chen2021transunet} & 85.08 & 77.02 & 81.87 & 63.16 & 94.08 & 75.62 & 87.23 & 55.86 & 77.49 &{-} &{-}\\
    Swin-UNet \cite{swinunet} & 90.66 & 79.61 & 83.28 & 66.53 & 94.29 & 76.60 & 85.47 & 56.58 & 79.13 &{-} &{-}\\
    MISSFormer \cite{missformer} & 91.92 & 82.00 & 85.21 & 68.65 & 94.41 & 80.81 & 86.99 & 65.67 & 81.96 & {-} &{-}\\
    Swin-UNETR ~\cite{SWINUNETR} 
        & 95.37 &  86.26 & 86.99 &  66.54 & 95.72 &  77.01  &  91.12 &   68.80  &   83.48 &{-} &{-}\\
    nnFormer~\cite{nnFormer} 
        & 90.51 & 86.25 & 86.57 & 70.17 &96.84 & \textbf{86.83}  & 92.04 &    \textbf{83.35} &   86.57 &{-} &{-}\\
       \hline
    MedNeXt-M-K3 \cite{roy2023mednext} 
        & 90.63 &  86.50 & 87.66 &  73.00 & 96.92 &  77.89  &  92.25 &  80.81 &  85.71 & {2.85e-04} & {0.999714}\\
    \textbf{MedNeXt-M-K3 (Ours)}
        & 92.65 & 87.42 & \textbf{87.73} & \textbf{73.25} & \textbf{96.93} &  78.55  &  \textbf{93.37}&  82.10  &   86.50& {2.54e-04}& {0.999781}\\
        \cline{2-10}
       & \multicolumn{9}{|c|}{p-value  = 7.55e-05 < 0.01}& &\\
        \hline
        MedNeXt-M-K5 \cite{roy2023mednext}
        &91.16 & 87.51 & 87.67 &  71.31 & 97.01 &  80.46 &  92.48 &   80.20  &  85.97& {2.28e-04} & {0.999772}\\
        MedNeXt-M-K5 (Ours)
        & 92.80 & 88.06 &87.70 & 71.85 & 96.89 &  81.55 &  93.12 &  81.63 &   86.70& {2.15e-04} & {0.999831}\\
         \cline{2-10}
     & \multicolumn{9}{|c|}{p-value  = 1.27e-04 < 0.01} & &\\
        \hline
        UNETR++~\cite{UNETR++} & \textbf{95.94} & 87.16 & 87.57 & 68.34 & 96.35 & 83.93 & 92.88 & 82.16  &   86.80 & {3.22e-04} & {0.999678}\\
         \textbf{UNETR++ (Ours)}  & 95.41 & \textbf{88.92} & 87.50 & 73.03 & 96.24 & 85.66 & 92.62 & 82.55  &   \textbf{87.74} &{2.88e-04} & {0.999712}\\
          \cline{2-10}
       & \multicolumn{9}{|c|}{p-value  = 4.80e-06 < 0.01} & &\\
        \bottomrule
        \end{tabular}
        }
\end{table*}

\subsection{Self Supervised Learning}
Self-supervised learning helps to reduce the dependency on extensive labeled datasets by leveraging the intrinsic information present within the data itself. Generally, in self-supervised pretraining, the models are trained to learn useful differentiable characteristics of the data via some pretext tasks such as predicting the angle of rotation, solving the jigsaw puzzle, etc. Several attempts including \cite{2022unetformer}, \cite{taleb20203d}, \cite{tang2022self}, \cite{zhang2023dive}, \cite{modelgenesis} have been made to design suitable self-supervised tasks for volumetric medical images which can capture the whole spatial context. 

Model genesis approach introduced in \cite{modelgenesis} formulated a single objective pretraining for CNN models based on image restoration proxy tasks. The first transformer-based self-supervised pretraining framework for 3D medical image analysis \cite{tang2022self} introduced a multi-objective pretext task combining rotation, masked volume inpainting, and contrastive coding. Unlike the model genesis pretraining approach, which involves using both encoder and decoder for a single objective pretraining, Swin UNETR pretraining is formulated as a multi-objective task with a separate loss function for each of the proxy tasks and makes use of only the encoder. However, Swin UNETR pretraining using five large-scale CT (Computed Tomography) datasets could not be used for the MRI (Magnetic Resonance Imaging) segmentation task due to the domain gap between CT and MRI images. Relying on self-supervised pretraining methods, which require large-scale datasets with the same domain characteristics, is not a practical solution for data-deficient medical domain applications. A self-supervised pretraining framework based on volumetric masking and reconstruction pretext task proposed in \cite{2022unetformer} also utilized the large cohort of 5050 images for pretraining the UNetFormer encoder. SwinMM pretraining approach introduced in \cite{wang2023swinmm} employs a multi-view encoder, a decoder with a cross-attention module, and follows a mutual learning paradigm to extract hidden multi-view information to generate precise segmentation masks.

Although self-supervised pretraining approaches introduced in  \cite{2022unetformer}, \cite{tang2022self}, \cite{wang2023swinmm} were proven to be effective for volumetric image segmentation, these methods heavily depend on large-scale medical datasets which consequently contributes to increased data and computational costs. The SOTA self-supervised pretraining methods for volumetric medical image segmentation, such as  Swin UNETR \cite{tang2022self} and SwinMM \cite{wang2023swinmm} rely on large-scale datasets posing limitations in generalizability for data-scarce medical image analysis tasks. Swin UNETR pretraining dataset includes 5050 CT scans from 5 public datasets namely LUNA16 \cite{setio2017validation}, TCIA Covid19 \cite{desai2020chest}, LIDC \cite{armato2011lung}, HNSCC \cite{grossberg2018imaging} and TCIA Colon \cite{johnson2009accuracy}. SwinMM network was pretrained using 5833 volumetric scans from 8 public datasets: AbdomenCT-1K \cite{ma2021abdomenct}, BTCV \cite{Synapse}, MSD \cite{MSD}, TCIA-Covid19, WORD \cite{luo2022word}, TCIA-Colon, LIDC, and HNSCC.

In this work, we propose a learnable weight initialization scheme that utilizes limited available training data to learn discriminative characteristics from the volumetric medical images, which can improve the model performance without the need for any additional data or higher computation costs. Our approach uniquely leverages volumetric self-supervised tasks on the same dataset for weight initialization and segmentation tasks in medical imaging, demonstrating efficiency and efficacy. 

\begin{figure*}[!t]
\setlength{\belowcaptionskip}{-2pt}
\setlength{\abovecaptionskip}{-2pt}
\setlength{\abovedisplayskip}{-2pt}
\setlength{\belowdisplayskip}{-2pt}
\begin{center}
\includegraphics[width=0.8\linewidth]{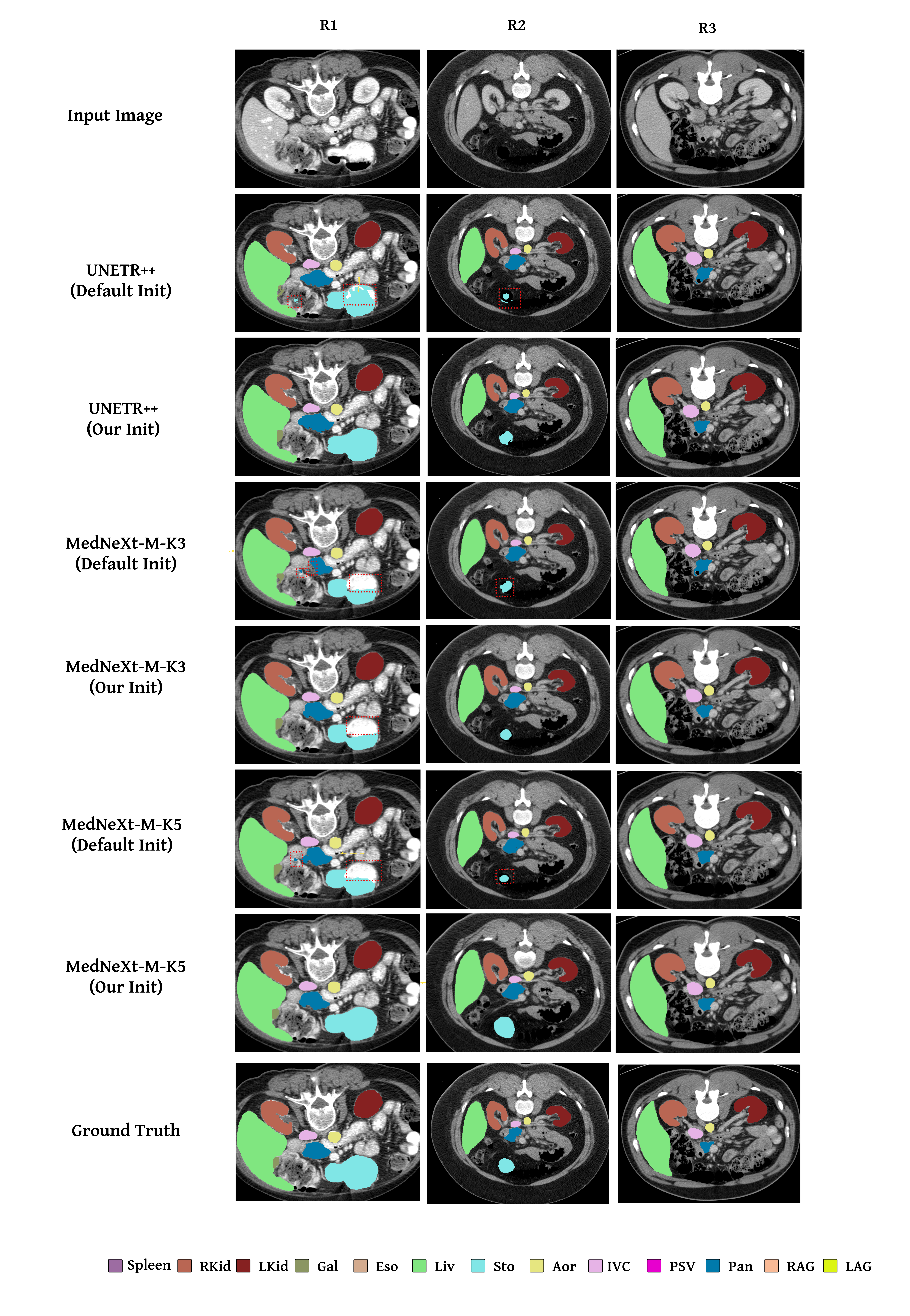}
\end{center}
   \caption{\textbf{ Qualitative results for Synapse dataset on SOTA segmentation networks:} The proposed data-dependent initialization
scheme, when integrated with different segmentation networks, improves the overall segmentation performance by accurately segmenting the organs and delineating organ boundaries. Organs are shown in the legend below the example images. Abbreviations are as follows: Spl: \em{spleen}, \em{RKid: right kidney}, \em{LKid: left kidney}, Gal:
    \em{gallbladder}, \em{Eso: esophagus}, Liv: \em{liver}, Sto: \em{stomach}, Aor: \em{aorta}, IVC: \em{inferior vena cava}, PSV: \em{portal and splenic veins}, Pan:
\em{pancreas}, RAG: \em{right adrenal gland}, and LAG: \em{left adrenal gland}. Best Viewed zoomed in.}
	\label{fig:Qual_synapse_SOTA}
\end{figure*}
\section{Method}
\subsection{Data Independent Weight Initialization}
\label{data_ind}

As discussed earlier, deep neural networks typically require a large amount of training data to achieve promising results. However, this is challenging in medical imaging tasks due to the scarcity of ample medical training data. Collecting and annotating medical images is a complex and expensive process. This becomes further problematic in the case of transformers-based medical segmentation approaches due to the lack of inductive biases, thereby requiring a large amount of training data. Most existing medical image segmentation methods \cite{SWINUNETR}, \cite{UNETR},\cite{UNETR++}, \cite{nnFormer}
address this issue by focusing on architectural improvements, such as integrating CNNs with ViTs to inherit the inductive biases, or using hierarchical structural representations. These hybrid CNN-transformers approaches typically strive to improve the locality of ViTs. However, they mostly utilize the standard \textit{data-independent} initialization schemes such as truncated normal, Xavier
\cite{glorot2010}, and Kaiming He \cite{he2015}, which do not explicitly take into account the volumetric characteristics of the medical segmentation data. For instance, the default weight initialization scheme in the UNETR framework \cite{UNETR} is 

\begin{flalign}
\label{eq: default initialization}
\hspace{-1 cm}
    \begin{cases}
    &\mathcal{U}(-\sqrt{\sigma}, \sqrt{\sigma}),   \sigma=\frac{1}{C * \prod_{i=0}^{2}ksize(i) };  \text{Convolutional Layers} \\ 
    &\mathcal{U}(-\sqrt{\sigma}, \sqrt{\sigma}), \sigma=\frac{1}{C}  ;  \hspace{5.3em}  \text{Linear Layers}
    \end{cases}
\end{flalign}

\noindent Where $\mathcal{U}$ is a continuous uniform distribution, $\sigma$ is the standard deviation, $C$ is the number of input channels, and $ksize$ is the kernel size at position $i$.

We observe that the choice of the initialization scheme plays an important role in network learning and can affect model convergence. For instance, Fig. \ref{fig:Diff_init} (left) shows that UNETR \cite{UNETR} converges to different solutions based on the model initialization. We can see a substantial decrease in performance for UNETR when initialized using the Kaiming approach, whereas the truncated normal approach yields improved outcomes compared to UNETR's default initialization scheme. Using data-independent initialization schemes can likely limit the performance since medical segmentation datasets have fewer samples when compared with large-scale natural image benchmarks. Therefore, the model may struggle to learn the representations effectively during the training when the number of training samples is relatively lower with respect to the network parameters. 
 
In this work, we propose a \textit{data-dependent learnable weight initialization} method that explicitly takes into account the volumetric nature of the medical data. Our approach induces structural and contextual consistency within encoder-decoder networks in the early stage of the training, leading to improved segmentation performance (e.g., fewer false positives and better delineation of segmentation boundaries) as shown in Fig. \ref{fig:Diff_init} (right). The useful prior knowledge about data-dependent biases learned by our approach provides a better starting point for model training, that leads to improved segmentation without utilizing additional data or increasing the computation costs.

\subsection{Learning Data-Dependent Weight Initialization}
\label{data_dep}


Our work focuses on designing a learnable weight initialization method for hybrid volumetric medical image segmentation frameworks. Consider a hybrid volumetric medical image segmentation network that consists of a ViT encoder $\mathcal{F}$ and a CNN-based decoder $\mathcal{G}$. The encoder converts 3D input patches into latent feature representations $\mathbf{Z}_{i}$ at multiple levels $i$. The output segmentation mask ($\hat{Y}$) is generated by combining encoder representations at multiple resolutions with the corresponding upsampled decoder representations. Given a 3D input volume $\mathbf{X} \in \mathbb{R}^{C \times{H} \times{W} \times{D}}$, where $C$, $H$, $W$, $D$ represents the number of channels, height, width, and depth of the image respectively, the latent feature representations generated by the encoder can be represented as: 

\begin{equation}
\label{eq: latent representations}
    \mathcal{F}(\mathbf{X}) = \mathbf{Z}_{i} 
    \in \mathbb{R}^{\frac{H}{P_{h_i}} \times \frac{W}{P_{w_i}} \times \frac{D}{P_{d_i}} \times E_{i}} 
    ;\;\;\;\   i = 1,2,\hdots, m
\end{equation}

\noindent where $E_i$ refers to the embedding size, $P_{h_i}$, $P_{w_i}$, and $P_{d_i}$, represent the patch resolution of the encoder representations at layer $i$ across height, width, and depth respectively, and $m$ is the total number of encoder layers connected to the decoder via skip connections.

Our proposed method consists of \textbf{(Step 1)} learnable weight initialization in which the model is trained on multi-objective self-supervised tasks to effectively capture the inherent data characteristics, followed by the \textbf{(Step 2)} supervised training for the volumetric segmentation task. 

Our method utilizes the same training dataset for both 
steps and is therefore beneficial for 
3D medical imaging segmentation tasks on standard benchmarks having limited data samples. Fig. \ref{fig:main_demo} presents an overview of our approach in a standard encoder-decoder 3D medical segmentation framework. We introduce a \emph{Transformation Module} during \textbf{Step 1} to generate masked and shuffled input volume and the encoder-decoder is trained to predict the correct order of medical scans while reconstructing the missing portions as described next.
 
\subsubsection*{Step I- Weight Initialization through Self-supervision}
Our approach injects structural and contextual consistency within the transformer architecture through the self-supervised objectives. To effectively capture the underlying patterns in the volumetric CT or MRI data, we transform the given input volume $\mathbf{X}$ using our proposed transformation module (Fig. \ref{fig:main_demo}). 

\noindent \textbf{Transformation Module:}
It rearranges the input volume across the depth and then partitions it to $\mathcal{B}$ non-overlapping equal-sized sub-volumes. For a given input volume of depth $D$, we define it as $\mathcal{B}=\frac{D}{P_{d_m}}$, where $P_{d_m}$ is the patch resolution at the encoder bottleneck ( $m^\text{th}$ level). We first rearrange the input $\mathbf{X}$ into sub-volumes such as, $\mathbf{X} = [ \mathbf{x}_1, \mathbf{x}_2, ..., \mathbf{x}_\mathcal{B}]$. These sub-volumes can be rearranged or shuffled in $\mathcal{B}!$ permutations. We randomly select a permutation sequence $\mathbf{O}$ out of them and shuffle the sub-volumes to generate $\mathbf{X'}$. We then apply random masking to the shuffled volume $\mathbf{X'}$ using a predefined masking ratio and patch size to obtain a masked and shuffled volume $\mathbf{X''}$. The masked and shuffled input volume is then processed by the model to learn structural and contextual consistency in the data.

\noindent\textbf{Structural Consistency through Order Prediction:} Our approach mines intrinsic anatomical information from volumetric scans to bring structural consistency to the transformer encoder by learning to predict the correct order of transformed shuffled input. 
This can be formulated as a classification task with $\mathcal{B}$ classes within the encoder latent space. We append a classifier head at the end of each encoder representation $\mathbf{Z}_i$ (Eq. \ref{eq: latent representations}). Then, we flatten and average the encoder representation at each layer $\mathbf{Z}_i, \{i=1,2,3, \hdots , m\}$ across the height and width dimension to obtain an intermediate embedding of size $\mathbb{R}^{\frac{D}{P_{d_m}} \times E_i}$. We forward pass these intermediate feature representations through their corresponding classifier to obtain the order prediction $\mathbf{t}_{i} \in \mathbb{R}^{\frac{D}{P_{d_m}} \times \mathcal{B}}$ (see Fig. \ref{fig:main_demo}). 

We define the structural consistency by predicting the correct order of shuffled input through cross-entropy loss between each output order prediction $\mathbf{t}_{i}$ and the ground truth permutation 
used for sub-volume shuffling. Our order prediction loss $\mathcal{L}_{Cls}$ is as follows:

\begin{equation}
\mathcal{L}_{Cls} = \sum_{i} \sum_{f=1}\left( -\sum_{k=1}^{\mathcal{B}}\mathbf{O}_{k,f} \log(\mathbf{t}_{i}) \right), 
\end{equation}
where $i=1,2,3,\hdots, m$ and $f=1, 2, ..., \frac{D}{P_{d_m}}$.

 \noindent Here, $\mathcal{B}$ represents the number of classes that corresponds to the number of sub-volumes. $\mathbf{O}_{k,f}$  corresponds to the ground truth order for sub-volume.

\noindent\textbf{Contextual Consistency through Voxel Reconstruction:} Our proposed initialization method utilizes 3-dimensional masking and reconstruction tasks to inject contextual consistency by learning the correspondence between the masked regions and their neighboring context. i.e, the model will be trained 
to reconstruct the masked volume $\mathbf{X''}$ at the decoder $\mathbf{Y'}= \mathcal{G}(\mathbf{X''})$. 
The reconstruction loss ($\mathcal{L}_{Rec}$) between the non-masked input $\mathbf{X'}$ and its corresponding reconstructed volume $\mathbf{Y'}$ is measured by voxel-wise mean square error calculated:

\begin{equation}
\mathcal{L}_{Rec} = \mathcal{L}_{MSE} (\mathbf{X'},\mathbf{Y'}) = \frac{1}{N}\sum_{n=1}^{N}{(\mathbf{X'}_n-\mathbf{Y'}_n)}^2,
\end{equation}

\noindent where $N$ represents the total number of voxels in the 3D volume. Our final self-supervised training loss $\mathcal{L}$ in the first step is computed as:
\begin{equation}
\mathcal{L} =\mathcal{L}_{Cls} + \mathcal{L}_{Rec}
\end{equation}
\subsubsection*{Step II- Training For Segmentation} 
During the second stage, the model is trained on the same training dataset for segmentation in a supervised fashion by utilizing a combined soft dice and cross-entropy loss \cite{milletari2016v}. The model weights learned from the first step are transferred to serve as a better initialization for the subsequent segmentation training task. Given an input volume $\mathbf{X}$ and its corresponding ground truth segmentation mask $\mathbf{Y}$, the model is trained in the second step with the following supervised objective:
\begin{equation}
    \mathcal{L} =\mathcal{L}_{Dice+CE}(\mathbf{Y},\hat{\mathbf{Y}})
\end{equation}
where $\hat{\mathbf{Y}}$ is the output segmentation mask  produced by the model and the loss function $\mathcal{L}_{Dice+CE}$ is the combination of cross-entropy and soft Dice: 

\begin{equation}
\label{eq: loss_func}
\begin{gathered}[b]
    \mathcal{L}_{Dice+CE}(\mathbf{Y},\hat{\mathbf{Y}}) = 1-\frac{2}{J}\sum_{j=1}^{J}\frac{\sum_{n=1}^{N} \mathbf{Y}_{n,j} \hat{\mathbf{Y}}_{n,j} }{\sum_{n=1}^{N}\mathbf{Y}^{2}_{n,j}+ \sum_{n=1}^{N}\hat{\mathbf{Y}}^{2}_{n,j}}\\
-\frac{1}{N}\sum_{n=1}^{N}\sum_{j=1}^{J} \mathbf{Y}_{n,j}\log \hat{\mathbf{Y}}_{n,j}
\end{gathered}
\end{equation}
\noindent where $J$ and $N$ represent the total number of class labels and voxels respectively. $\hat{\mathbf{Y}}_{n,j}$ and $\mathbf{Y}_{n,j}$ denotes the model output and corresponding ground truth probabilities for a class $j$ at a specific voxel $n$. 

\begin{table}
\centering
\caption{\textbf{Baseline Comparison - Lung Dataset:} Our approach helps to reduce the False positives and better delineate the organ boundaries as indicated  by the improvements in terms of Dice score (DSC), False Positive
Rate (FPR) and True Negative Rate (TNR). }
\label{table: base_lung}
\begin{tabular}{l c c c}
\toprule
Model &  DSC  & {FPR} & {TNR} \\
\midrule
UNETR ~\cite{UNETR} & 69.11 &  {3.83e-05} &  {0.999961}\\
\addlinespace
 \textbf{UNETR(Ours)} &  \textbf{70.28} &  {3.79e-05}  & {0.999962}\\
 \midrule
\multicolumn{4}{c}{p-value = \small{$1.20\mathrm{e}{-05}$} < 0.01 }    \\
\bottomrule 
\end{tabular}
\end{table}

\begin{table}
\centering
\caption{\textbf{SOTA Comparison - Lung Dataset:} Integrating our proposed weight intialization approach helps to improve the segmentation performance in terms of Dice score (DSC), False Positive Rate (FPR) and True Negative Rate (TNR).}
\label{SOTA:lung}
\setlength{\tabcolsep}{3.5pt}
\begin{tabular} { l c c c } 
\toprule
Method & DSC &  {FPR} & {TNR} \\
\midrule
nnUNet  ~\cite{nnUnet} &  74.31 &{-} & {-} \\
Swin UNETR  ~\cite{SWINUNETR} &75.55 & {-} &{-}\\
nnFormer ~\cite{nnFormer} & 77.95 & {-} & {-}\\
\midrule
MedNeXt-M-K3 \cite{roy2023mednext} & 80.54 & 1.48e-05 & 0.999985\\
MedNeXt-M-K3 (Ours)& 81.26 & 1.29e-05 & 0.999989 \\
\midrule
\multicolumn{4}{c}{p-value = 1.40e-04 < 0.01} \\
\midrule
MedNeXt-M-K5 \cite{roy2023mednext} &  {79.51}& {1.77e-05} & {0.999982}\\
{MedNeXt-M-K5 (Ours) } &{80.60}& {1.63e-05} &{0.999987}\\
\midrule
\multicolumn{4}{c}{p-value = 8.14e-05 < 0.01} \\
\midrule
 UNETR++ ~\cite{UNETR++} & 80.68 & {1.82e-05} & {0.999943}\\
 \textbf{UNETR++ (Ours)} & \textbf{81.69} & {1.39e-05} & {0.999986}\\
\midrule
\multicolumn{4}{c}{p-value = 6.98e-06 < 0.01} \\
\bottomrule 
\end{tabular}
\end{table}
\subsection{Generalizability}
In contrast to the existing practice of initializing the models using large-scale natural image dataset (ImageNet) pre-trained weights, or using generic initialization schemes adopted from mainstream computer vision, we transfer the weights learned from the first step to initialize the model training in the second step. The self-supervised inductive biases learned during the initialization stage will serve as an effective weight initialization scheme for the subsequent segmentation training task. 

Our proposed data-dependent weight initialization approach is complementary and can be integrated into any volumetric segmentation model to provide a better starting point for model training by learning initial model weights via our proposed self-supervised tasks without modifying the architecture or loss functions. We show that our data-dependent weight initialization scheme performs seamlessly well for both fixed-size representation models like UNETR \cite{UNETR} and hierarchical representation models like  Swin-UNETR \cite{SWINUNETR} UNETR++ \cite{UNETR++} and MedNeXt \cite{roy2023mednext}.
\section{Results and Analysis}
\subsection{Datasets}
\noindent We validate the effectiveness of our proposed approach on the following two datasets: 

\noindent \textbf{Synapse for Multi-organ CT Segmentation:}
Synapse  \cite{Synapse} is a CT dataset that consists of abdomen scans of 30 subjects with 8 organs: \emph{spleen}, \emph{right kidney}, \emph{left kidney}, \emph{gallbladder}, \emph{liver}, \emph{stomach}, \emph{aorta} and \emph{pancreas}. Each CT scan has around 80 to 220 slices with $512 \times 512$ pixels. Following the previous approaches, we utilized the data split provided in \cite{chen2021transunet} to train our models on 18 training samples and evaluated them using 12 validation samples. 

\noindent \textbf{Decathlon Lung Dataset:}
Lung dataset from Medical Segmentation Decathlon (MSD) \cite{MSD} for lung cancer segmentation consists of CT volumes of 63 subjects. Lung cancer segmentation is formulated as a binary segmentation task (background or lung cancer). We split the data into 80:20 ratio for training and validation for the experiments.

\begin{figure*}[!t]
\centering
 \begin{minipage}[b]{0.3\linewidth}
\centering
\includegraphics[width=\linewidth]{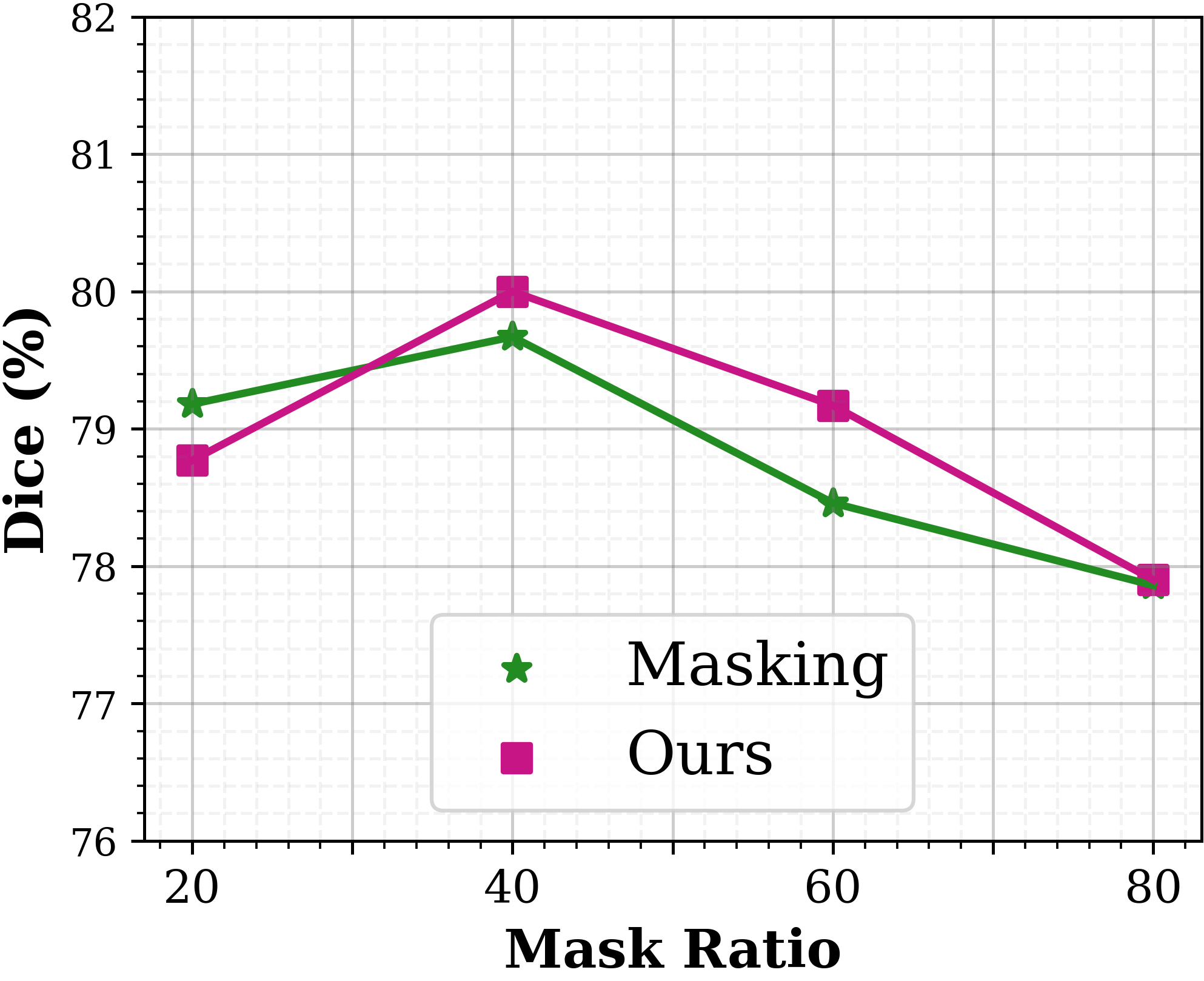}
\vspace{3pt}
(a) 
\end{minipage}
 \begin{minipage}[b]{.3\linewidth}
\centering
\includegraphics[width=\linewidth]{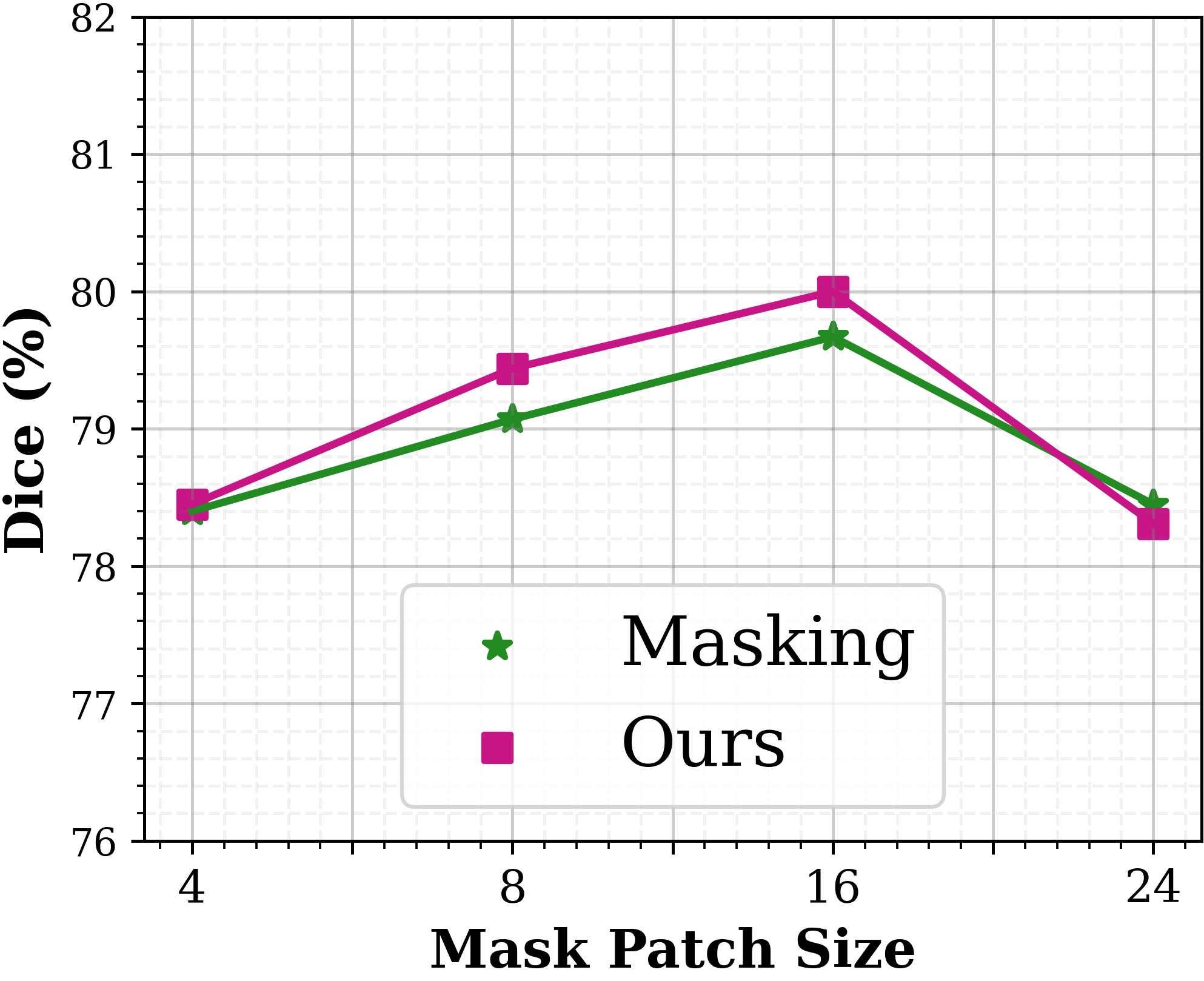} 
\vspace{3pt}
(b) 
 \end{minipage} 
\begin{minipage}[b]{0.3\linewidth}
\centering
 \includegraphics[width= \linewidth]{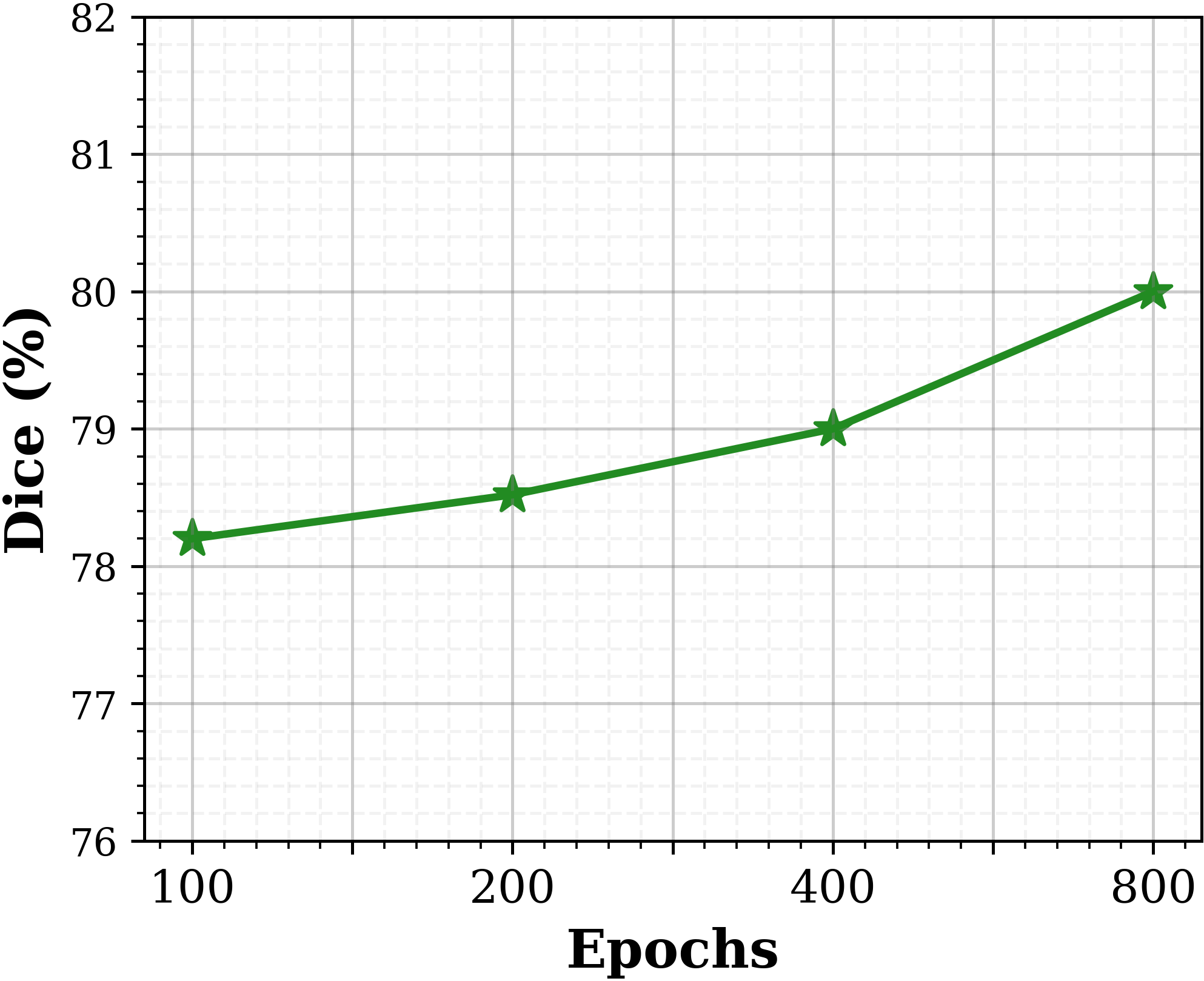}
\vspace{3pt}
(c) 
\end{minipage}
\caption{\textbf{Effect of masking ratio (a) and mask patch size (b):} Moderate masking with masking ratio around 40\% and mask patch size of ($16 \times 16 \times 16$) during the initialization step (step-1) yields the optimal results for UNETR on synapse dataset. \textbf{Effect of increasing the training epochs for initialization (c):} Training on our proposed approach on initialization for longer epochs improves the overall segmentation performance.}
\label{fig:combined_mask}
\end{figure*}

\subsection{Implementation Details} 
\noindent We implemented our approach in Pytorch and Monai. For a fair comparison, we used the same input size and pre-processing steps of UNETR and UNETR++ for our experiments. We train all the models using a single A100 40GB GPU and use a sliding window approach with an overlap of $0.5$ for inference and report the model performance in percentage Dice score (Dice \%). All the results are reported based on single model accuracy without any ensemble or additional data. 
\subsection{Baseline Comparison}
\noindent Table \ref{table:UNETR synapse} 
and Table \ref{table: base_lung}
illustrate the impact of our proposed data-dependent initialization approach on the UNETR performance when trained on multi-organ Synapse and Decathlon Lung datasets. For a fair comparison, all models are trained on 3D input volumes of size $ 96 \times 96 \times 96 $ following the UNETR training framework. Our approach achieves an absolute gain of 2.56\% over the baseline UNETR for the Synapse dataset with significant improvement in the segmentation results of smaller organs such as the aorta, gallbladder, and pancreas. 

\begin{figure} 
\centering
\includegraphics[width=\columnwidth]{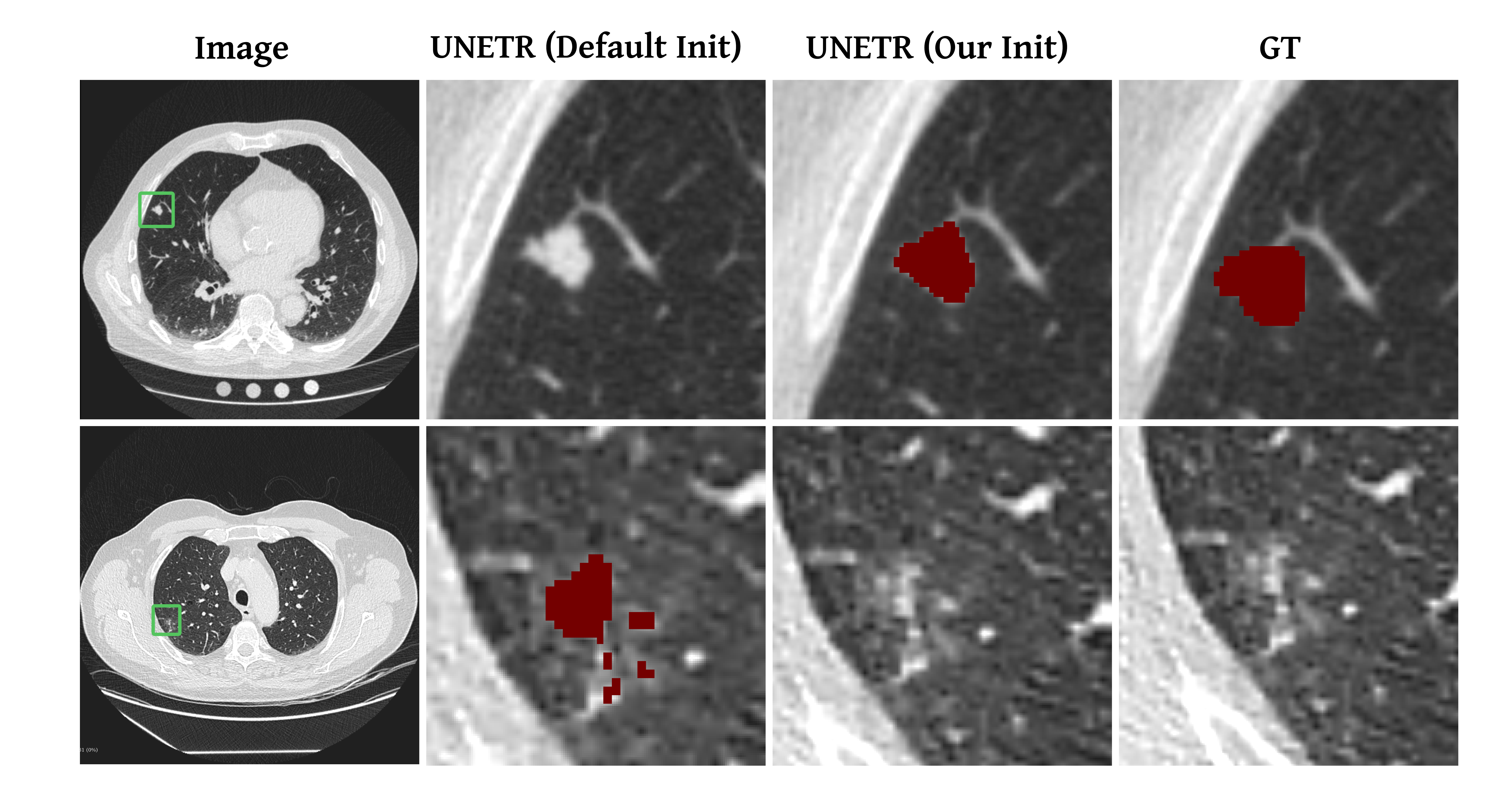}
   \caption{\textbf{Qualitative results (Lung) on UNETR:} Columns 2-4 show the enlarged views of the segmented areas marked in
a green box in column 1. Integrating our proposed learnable initialization approach is beneficial in learning the structural and contextual cues from the training data, which helps in reducing the cases of miss classification (false negatives (row 1) and false positives (row 2)).}
	\label{fig:Qualitative_lung_UNETR}
\end{figure}

For the decathlon-lung dataset, by integrating our proposed weight initialization approach, the lung-cancer segmentation result improved by 1.17\%,  compared to the UNETR baseline as shown in Table \ref{table: base_lung}. It is clear from Fig. \ref{fig:Qualitative_lung_UNETR} that our approach improves lung cancer segmentation by reducing the instances of miss classification.

\begin{table}\normalsize
\centering
\caption{Network configuration of UNETR++ and MedNeXt}
\label{Table: config}
\begin{tabular}{l |c |c}
\hline
Attribute & Synapse  &  Lungs \\
\hline
Spacing & [0.76, 0.76, 3] & [1.52, 1.52, 6.35]\\
Crop size & $(128 \times 128 \times 64)$ & $(192 \times 192 \times 32 )$\\
 Batch size & 2 & 2\\
\hline 
\end{tabular}
\end{table}
\subsection{State-of-the-Art Comparison}

We integrate our approach with state-of-the-art methods such as MedNeXt and UNETR++ to enhance their segmentation performance on synapse and lung datasets. The organ-wise results in Table \ref{table: SOTA_synapse} and Table \ref{SOTA:lung} reveal that, unlike many existing approaches that fail to achieve satisfactory results across different organs, our approach excels by consistently delivering high performance for all organs. As depicted in Fig. \ref{fig:Qual_synapse_SOTA}, our approach improves upon the state-of-the-art UNETR++ on synapse by precisely delineating organ boundaries. For a fair comparison, all the experiments on UNETR++ and MedNeXt were performed using the same network configuration as shown in Table \ref{Table: config}. To integrate our proposed method, the models were trained on the initialization step for 200 epochs with a learning rate of 1e-4, prior to the 1000 epochs of training for segmentation with a learning rate of 1e-2.

\begin{figure} [t]
\centering
\includegraphics[width=\columnwidth]{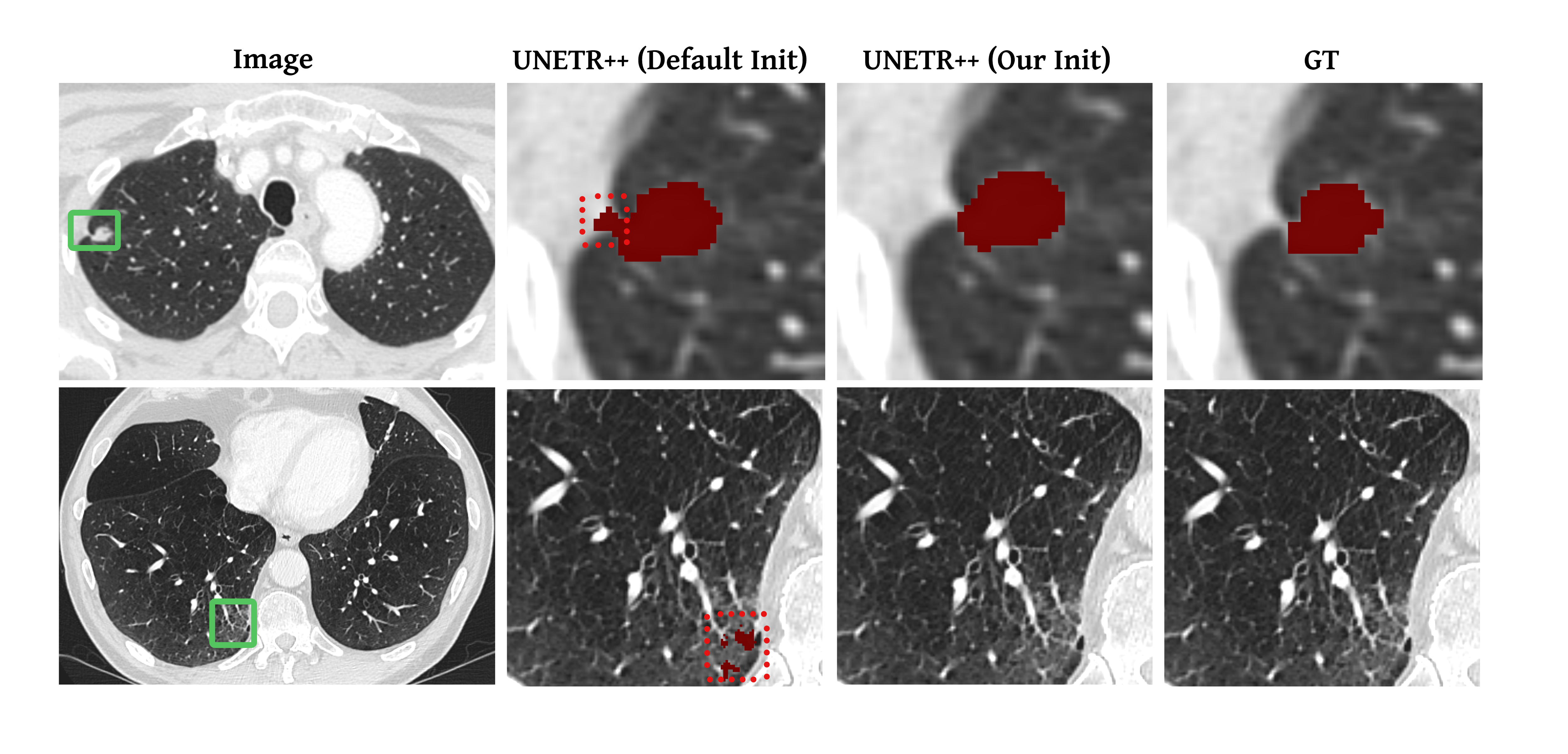}
   \caption{\textbf{Qualitative results (Lung) on UNETR++ :} Columns 2-4 show the enlarged views of the segmented areas marked in a green box in column 1. Our approach reduces the false positives (marked in red dashed box). Best Viewed zoomed in.}
	\label{fig:Qualitative_Lungs_SOTA}
\end{figure}

The qualitative comparison of Lung dataset segmentation results given in Fig. \ref{fig:Qualitative_Lungs_SOTA}
indicate that our approach helps in reducing the false positives for lung cancer segmentation and thereby improves the Decathlon-Lung state-of-the-art results on UNETR++ by 1.01\% as shown in Table \ref{SOTA:lung}.
\subsection{Statistical Significance}
We conducted independent two-sample t-tests to compare the average Dice scores between the baseline model and our corresponding weight-initialized model (referred to as `Ours'). The null hypothesis assumes that our approach provides no advantage over the baseline. On both the Synapse and Lung datasets, our proposed weight initialization-based models consistently yielded p-values less than 0.01 when compared with their respective baseline models, as demonstrated in Tables \ref{table:UNETR synapse}, \ref{table: SOTA_synapse}, \ref{SOTA:lung}, and \ref{table: base_lung}. These results strongly suggest the superiority of our approach over the baseline.

\begin{table*}[!t]\normalsize
\centering
\caption{Effect of different self-supervised objectives on UNETR performance.}
    \label{table: diff_pretrain}
       \centering
        \scalebox{0.86}{
        \begin{tabular}{c c c c c c c c c c}
        \toprule
           Weight Initialization  & Spleen &  R.kidney &   L.kidney & G.bladder  & Liver  & Stomach & Aorta &  Pancreas &  \textbf{Average}\\   
                 \midrule
     Default ~\cite{UNETR}& \textbf{88.58} & 80.03& 78.87 & 62.51 & 95.45 & 74.44 & 84.79 & 52.70  &   77.17 \\
    \cmidrule(lr){2-10}
    masking \cite{he2022masked} 
        & 85.75 & 82.26  & 84.50 & 59.84  & 95.61 & 72.37 & \textbf{88.14}  &  60.17   & 78.58  \\
       \cmidrule(lr){2-10}
    Tube masking ~\cite{feichtenhofer2022masked} 
        & 87.54 & 82.48 & 83.90 & 58.66 &  95.36 & 73.48 & 86.55 &  58.25 & 78.28   \\
      \cmidrule(lr){2-10}
      
        Order prediction (Ours) & 88.22 & \textbf{83.42}  & 85.03 &  62.08 & 95.36 &  70.56 & 86.86  & \textbf{60.24} & 78.97  \\
\cmidrule(lr){2-10}
       \textbf{Order prediction + masking (Ours)}  & 86.72 & 82.86 & \textbf{85.41} & \textbf{65.15} & \textbf{95.56} & \textbf{75.23} & 88.07  & 58.85 & \textbf{79.73} \\ 
        \bottomrule
        \end{tabular}
       }
\end{table*}

\subsection{Ablation Study}
\label{ablation}

\noindent \textbf{Masking Ratio:} We evaluated the effect of the masking ratio and mask patch size used in the proposed self-supervised task and we could observe that very small and large patch sizes for masking as well as the masking ratio are not suitable for effective performance (Fig. \ref{fig:combined_mask}(a), (b)).
For an input volume size of $ 96 \times 96 \times 96 $, the experimental results on multi-organ synapse dataset, a masking ratio of 40\%, and mask patch size of $16 \times 16 \times 16$ during the initialization step results in the best downstream segmentation performance. The optimal masking ratio and mask patch size may vary based on the network and dataset characteristics since medical images have different modalities and intensity ranges. Hence, they can be considered as hyperparameters which can be tuned using a held-out validation set .

\noindent \textbf{Effect of training epochs:}  
 We studied the effect of varying the training period for the initializatin step (step-1) on UNETR performance for synapse dataset and the results are illustrated in Fig. \ref{fig:combined_mask} (c). We observed that large number of epochs during step-1 of our approach  helps to better capture volumetric data characteristics which in return further increases the performance. We set the number of epochs during Step-1 to 800.

Table \ref{Table: TRAIN_epochs} shows that integrating our data-specific initialization to UNETR without increasing the total training epochs (800 epochs of step-1 and 4200 epochs of step-2) can also improve results while increasing the training epochs (5800 epochs) of UNETR without our initialization does not lead to notable improvements.

\begin{table}[!htbp]
    \centering
    \caption{Efficiency in terms of total training epochs.}
    \label{Table: TRAIN_epochs}
    \begin{tabular}{c | c c | c c}
        \toprule
        Method & \multicolumn{2}{c}{\textbf{UNETR}} & \multicolumn{2}{c}{\textbf{UNETR (Ours)}} \\
      \cmidrule(lr){1-1}   \cmidrule(lr){2-3} \cmidrule(lr){4-5}
        Epochs & \multicolumn{1}{c}{5000} &  \multicolumn{1}{c}{5800} &  \multicolumn{1}{c}{800+4200} & \multicolumn{1}{c}{800+5000} \\
       \cmidrule(lr){1-5} 
        DSC(\%) & \multicolumn{1}{c}{77.17} & \multicolumn{1}{c}{77.46} & \multicolumn{1}{c}{79.20} & \multicolumn{1}{c}{79.73} \\
        \bottomrule
    \end{tabular}
\end{table}

\begin{figure}[t]
\centerline{\includegraphics[width=\columnwidth]{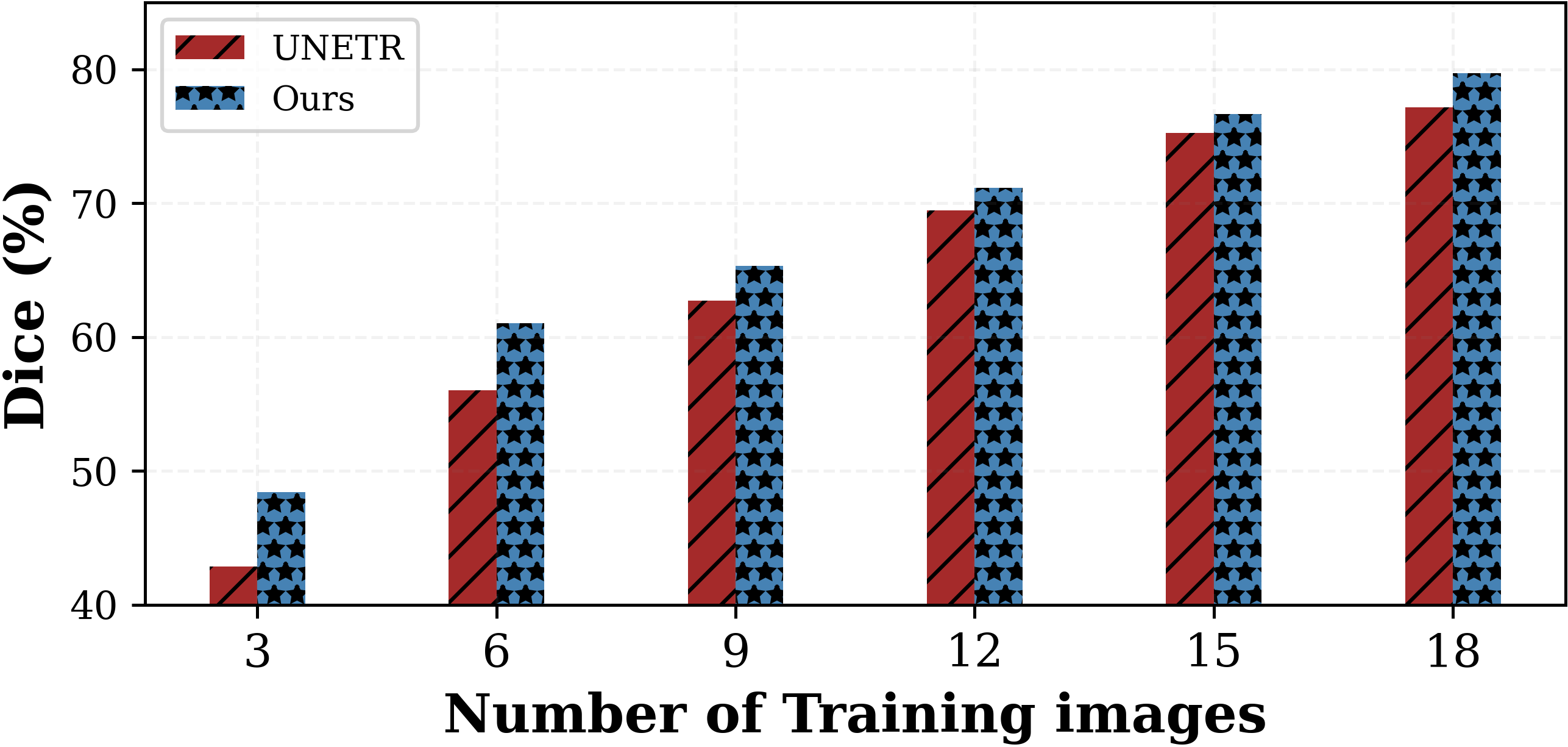}}
 \caption{\textbf{Effect of Training data size:} Our approach outperforms the UNETR baseline by high margin when trained with few data examples.}
\label{fig: TRAIN_DATA}
\end{figure}

\noindent \textbf{Effect of training data size:} We evaluated how varying the amount of data available during training impacts the UNETR performance for the synapse dataset. The results as shown in Fig. \ref{fig: TRAIN_DATA} demonstrate that our approach consistently performs better than the UNETR baseline, especially with fewer training examples. The ability to enhance results in such data-constrained scenarios is particularly valuable, as it helps to overcome challenges associated with insufficient data and improves the overall applicability and effectiveness of the segmentation approach. 

\noindent \textbf {Comparing different self supervised objectives:}
The effect of different self-supervised objectives on UNETR performance is discussed in Table \ref{table: diff_pretrain}. Our proposed combination of multi-objective self-supervised tasks performs better than existing masking-based self-supervised pretraining approaches \cite{he2022masked,feichtenhofer2022masked} and shuffling order prediction alone. These results demonstrate that our approach combines the advantages of both masked image reconstruction and shuffled order prediction effectively by capturing structural as well as contextual consistencies in the available training data.

\noindent \textbf{Effect of intermediate encoder representations:}
We studied the effect of utilizing intermediate encoder representations for order prediction during the initialization step (step-1), and the results are shown in Table \ref{table: cls_heads}. Here, $t_i$ represents the order predictions generated using intermediate encoder representation at layer $i$. We observe that incorporating the order predictions from multiple resolutions captures details at various levels of granularity,  encouraging extraction of fine-grained details and accurate delineation of boundaries, thereby improving the overall segmentation performance. 
\begin{table}\normalsize
\centering
\caption{Effect of intermediate encoder representations}
\label{table: cls_heads}
\begin{tabular}{l l l l| c c c c}
\hline
Classifier & & & & & & DSC (\%) &\\
\hline
t4  & & & & & & 77.87 &\\
t3 + t4 & & & & & & 78.60 & \\
t2 + t3 + t4  & & & & & & 79.10  & \\
 \textbf{t1+t2+t3+t4 (Ours)} & & & & & & \textbf{79.73} &\\
\hline 
\end{tabular}
\end{table}

\noindent \textbf{Comparison with Swin-UNETR large scale pretraining:} We compared our proposed initialization approach against the self-supervised pretraining approach introduced in \cite{tang2022self}. The Swin-UNETR pretraining framework involves pretraining the Swin-UNETR encoder using a large dataset cohort of 5050 CT images on multi-objective self-supervised tasks to learn robust feature representations to improve downstream segmentation, followed by finetuning of the whole model using the downstream segmentation dataset in a supervised manner. We used the publicly available Swin-UNETR pre-trained model weights and finetuned the model for synapse and BTCV datasets.  As shown in Table \ref{Table: swin_pre}, our proposed learnable weight initialization framework by utilizing only the available training data during both the self-supervised initialization and supervised segmentation training steps, we could achieve promising results comparable to the single model results of pretrained Swin-UNETR model in terms of average dice score. For a fair comparison of the self-supervised tasks, we also pretrained the Swin-UNETR encoder using only the available training images and then fine-tuned the model using the same training images. The results indicate that by using the available training data, our initialization approach outperforms the Swin-UNETR pretext tasks.

\begin{table} 
\footnotesize
\centering
\caption{Swin-UNETR single model results with different pretraining settings and our proposed weight initialization approach}
\label{Table: swin_pre}
\begin{tabular}{c|c|c|c|c}
\toprule
Method & Pretrained & Pretraining &  Synapse & BTCV\\
& & dataset & & \\
\midrule
 ~\cite{SWINUNETR} & $\times$ & - & 79.71 &  80.51\\
\midrule
  &  &  LUNA16 \cite{setio2017validation}, &  &  \\
&  & TCIA Covid19 \cite{desai2020chest}, &  &  \\
~\cite{tang2022self} & $\checkmark$ &  LiDC \cite{armato2011lung}, & 81.08 & 81.59 \\
&  &  HNSCC \cite{grossberg2018imaging}, &  &  \\
&  &  TCIA Colon \cite{johnson2009accuracy} &  &  \\
\midrule
 ~\cite{tang2022self} &  $\checkmark$ & Synapse/BTCV & 80.80 &  80.92\\
\midrule
Ours &  $\times$ & - & 81.41 & 81.60 \\ 
\bottomrule 
\end{tabular}
\end{table}

\subsection{Limitations of the proposed approach:}  Our approach incorporates an additional step of learnable weight initialization through self-supervised pretraining, introducing an associated computational cost. While this adds to the training process, it is essential to emphasize the substantial benefit as it improves the downstream segmentation results by learning the structural and contextual dependencies in the data. However, the cost for our proposed weight initialization step is significantly less when compared to the large-scale pretraining cost associated with pretraining techniques such as \cite{tang2022self} and \cite{wang2023swinmm} which relies on additional large-scale pretraining datasets.
\section{Conclusion}
In this work, we introduce a data-dependent weight initialization scheme that is designed to capture the volumetric data characteristics effectively in order to improve the downstream segmentation task. We propose to first train the model on tailored multi-objective self-supervised tasks to learn the contextual and structural consistency from the limited training data. The trained model weights will then be utilized to initialize the supervised training for segmentation. We demonstrate that our approach is complementary and can be easily integrated into any hybrid segmentation model to improve performance. 

\printcredits

\bibliographystyle{abbrv}
\bibliography{main}

\begin{thebibliography}{10}

\bibitem{armato2011lung}
S.~G. Armato~III, G.~McLennan, L.~Bidaut, M.~F. McNitt-Gray, C.~R. Meyer, A.~P. Reeves, B.~Zhao, D.~R. Aberle, C.~I. Henschke, E.~A. Hoffman, et~al.
\newblock The lung image database consortium (lidc) and image database resource initiative (idri): a completed reference database of lung nodules on ct scans.
\newblock {\em Medical physics}, 38(2):915--931, 2011.

\bibitem{awais2023foundational}
M.~Awais, M.~Naseer, S.~Khan, R.~M. Anwer, H.~Cholakkal, M.~Shah, M.-H. Yang, and F.~S. Khan.
\newblock Foundational models defining a new era in vision: A survey and outlook.
\newblock {\em arXiv preprint arXiv:2307.13721}, 2023.

\bibitem{cai2020dense}
S.~Cai, Y.~Tian, H.~Lui, H.~Zeng, Y.~Wu, and G.~Chen.
\newblock Dense-unet: a novel multiphoton in vivo cellular image segmentation model based on a convolutional neural network.
\newblock {\em Quantitative imaging in medicine and surgery}, 10(6):1275, 2020.

\bibitem{swinunet}
H.~Cao, Y.~Wang, J.~Chen, D.~Jiang, X.~Zhang, Q.~Tian, and M.~Wang.
\newblock Swin-unet: Unet-like pure transformer for medical image segmentation.
\newblock In {\em European conference on computer vision}, pages 205--218. Springer, 2022.

\bibitem{chen2021transunet}
J.~Chen, Y.~Lu, Q.~Yu, X.~Luo, E.~Adeli, Y.~Wang, L.~Lu, A.~L. Yuille, and Y.~Zhou.
\newblock Transunet: Transformers make strong encoders for medical image segmentation.
\newblock {\em arXiv preprint arXiv:2102.04306}, 2021.

\bibitem{Deeplab}
L.-C. Chen, G.~Papandreou, I.~Kokkinos, K.~Murphy, and A.~L. Yuille.
\newblock Deeplab: Semantic image segmentation with deep convolutional nets, atrous convolution, and fully connected crfs.
\newblock {\em IEEE transactions on pattern analysis and machine intelligence}, 40(4):834--848, 2017.

\bibitem{3dunet}
{\"O}.~{\c{C}}i{\c{c}}ek, A.~Abdulkadir, S.~S. Lienkamp, T.~Brox, and O.~Ronneberger.
\newblock 3d u-net: learning dense volumetric segmentation from sparse annotation.
\newblock In {\em Medical Image Computing and Computer-Assisted Intervention--MICCAI 2016: 19th International Conference, Athens, Greece, October 17-21, 2016, Proceedings, Part II 19}, pages 424--432. Springer, 2016.

\bibitem{desai2020chest}
S.~Desai, A.~Baghal, T.~Wongsurawat, P.~Jenjaroenpun, T.~Powell, S.~Al-Shukri, K.~Gates, P.~Farmer, M.~Rutherford, G.~Blake, et~al.
\newblock Chest imaging representing a covid-19 positive rural us population.
\newblock {\em Scientific data}, 7(1):414, 2020.

\bibitem{ViTs}
A.~Dosovitskiy, L.~Beyer, A.~Kolesnikov, D.~Weissenborn, X.~Zhai, T.~Unterthiner, M.~Dehghani, M.~Minderer, G.~Heigold, S.~Gelly, et~al.
\newblock An image is worth 16x16 words: Transformers for image recognition at scale.
\newblock {\em arXiv preprint arXiv:2010.11929}, 2020.

\bibitem{feichtenhofer2022masked}
C.~Feichtenhofer, Y.~Li, K.~He, et~al.
\newblock Masked autoencoders as spatiotemporal learners.
\newblock {\em Advances in neural information processing systems}, 35:35946--35958, 2022.

\bibitem{glorot2010}
X.~Glorot and Y.~Bengio.
\newblock Understanding the difficulty of training deep feedforward neural networks.
\newblock In {\em Proceedings of the thirteenth international conference on artificial intelligence and statistics}, pages 249--256. JMLR Workshop and Conference Proceedings, 2010.

\bibitem{grossberg2018imaging}
A.~J. Grossberg, A.~S. Mohamed, H.~Elhalawani, W.~C. Bennett, K.~E. Smith, T.~S. Nolan, B.~Williams, S.~Chamchod, J.~Heukelom, M.~E. Kantor, et~al.
\newblock Imaging and clinical data archive for head and neck squamous cell carcinoma patients treated with radiotherapy.
\newblock {\em Scientific data}, 5(1):1--10, 2018.

\bibitem{SWINUNETR}
A.~Hatamizadeh, V.~Nath, Y.~Tang, D.~Yang, H.~R. Roth, and D.~Xu.
\newblock Swin unetr: Swin transformers for semantic segmentation of brain tumors in mri images.
\newblock In {\em International MICCAI Brainlesion Workshop}, pages 272--284. Springer, 2021.

\bibitem{UNETR}
A.~Hatamizadeh, Y.~Tang, V.~Nath, D.~Yang, A.~Myronenko, B.~Landman, H.~R. Roth, and D.~Xu.
\newblock Unetr: Transformers for 3d medical image segmentation.
\newblock In {\em Proceedings of the IEEE/CVF winter conference on applications of computer vision}, pages 574--584, 2022.

\bibitem{2022unetformer}
A.~Hatamizadeh, Z.~Xu, D.~Yang, W.~Li, H.~Roth, and D.~Xu.
\newblock Unetformer: A unified vision transformer model and pre-training framework for 3d medical image segmentation.
\newblock {\em arXiv preprint arXiv:2204.00631}, 2022.

\bibitem{he2022masked}
K.~He, X.~Chen, S.~Xie, Y.~Li, P.~Doll{\'a}r, and R.~Girshick.
\newblock Masked autoencoders are scalable vision learners.
\newblock In {\em Proceedings of the IEEE/CVF Conference on Computer Vision and Pattern Recognition}, pages 16000--16009, 2022.

\bibitem{he2015}
K.~He, X.~Zhang, S.~Ren, and J.~Sun.
\newblock Delving deep into rectifiers: Surpassing human-level performance on imagenet classification.
\newblock In {\em Proceedings of the IEEE international conference on computer vision}, pages 1026--1034, 2015.

\bibitem{huang2020unet}
H.~Huang, L.~Lin, R.~Tong, H.~Hu, Q.~Zhang, Y.~Iwamoto, X.~Han, Y.-W. Chen, and J.~Wu.
\newblock Unet 3+: A full-scale connected unet for medical image segmentation.
\newblock In {\em ICASSP 2020-2020 IEEE International Conference on Acoustics, Speech and Signal Processing (ICASSP)}, pages 1055--1059. IEEE, 2020.

\bibitem{missformer}
X.~Huang, Z.~Deng, D.~Li, and X.~Yuan.
\newblock Missformer: An effective medical image segmentation transformer.
\newblock {\em arXiv preprint arXiv:2109.07162}, 2021.

\bibitem{nnUnet}
F.~Isensee, P.~F. Jaeger, S.~A. Kohl, J.~Petersen, and K.~H. Maier-Hein.
\newblock nnu-net: a self-configuring method for deep learning-based biomedical image segmentation.
\newblock {\em Nature methods}, 18(2):203--211, 2021.

\bibitem{johnson2009accuracy}
C.~D. Johnson, M.-H. Chen, A.~Y. Toledano, J.~P. Heiken, A.~Dachman, M.~D. Kuo, C.~O. Menias, B.~Siewert, J.~I. Cheema, R.~G. Obregon, et~al.
\newblock Accuracy of ct colonography for detection of large adenomas and cancers.
\newblock {\em Obstetrical \& Gynecological Survey}, 64(1):35--37, 2009.

\bibitem{karimi2021convolution}
D.~Karimi, S.~D. Vasylechko, and A.~Gholipour.
\newblock Convolution-free medical image segmentation using transformers.
\newblock In {\em Medical Image Computing and Computer Assisted Intervention--MICCAI 2021: 24th International Conference, Strasbourg, France, September 27--October 1, 2021, Proceedings, Part I 24}, pages 78--88. Springer, 2021.

\bibitem{khan2022transformers}
S.~Khan, M.~Naseer, M.~Hayat, S.~W. Zamir, F.~S. Khan, and M.~Shah.
\newblock Transformers in vision: A survey.
\newblock {\em ACM computing surveys (CSUR)}, 54(10s):1--41, 2022.

\bibitem{lahoud20223d}
J.~Lahoud, J.~Cao, F.~S. Khan, H.~Cholakkal, R.~M. Anwer, S.~Khan, and M.-H. Yang.
\newblock 3d vision with transformers: A survey.
\newblock {\em arXiv preprint arXiv:2208.04309}, 2022.

\bibitem{Synapse}
B.~Landman, Z.~Xu, J.~Igelsias, M.~Styner, T.~Langerak, and A.~Klein.
\newblock Miccai multi-atlas labeling beyond the cranial vault--workshop and challenge.
\newblock In {\em Proc. MICCAI Multi-Atlas Labeling Beyond Cranial Vault—Workshop Challenge}, volume~5, page~12, 2015.

\bibitem{lee20223d}
H.~H. Lee, S.~Bao, Y.~Huo, and B.~A. Landman.
\newblock 3d ux-net: A large kernel volumetric convnet modernizing hierarchical transformer for medical image segmentation.
\newblock {\em arXiv preprint arXiv:2209.15076}, 2022.

\bibitem{li2018visualizing}
H.~Li, Z.~Xu, G.~Taylor, C.~Studer, and T.~Goldstein.
\newblock Visualizing the loss landscape of neural nets.
\newblock {\em Advances in neural information processing systems}, 31, 2018.

\bibitem{liu2022convnet}
Z.~Liu, H.~Mao, C.-Y. Wu, C.~Feichtenhofer, T.~Darrell, and S.~Xie.
\newblock A convnet for the 2020s.
\newblock In {\em Proceedings of the IEEE/CVF conference on computer vision and pattern recognition}, pages 11976--11986, 2022.

\bibitem{FCN}
J.~Long, E.~Shelhamer, and T.~Darrell.
\newblock Fully convolutional networks for semantic segmentation.
\newblock In {\em Proceedings of the IEEE conference on computer vision and pattern recognition}, pages 3431--3440, 2015.

\bibitem{ma2021abdomenct}
J.~Ma, Y.~Zhang, S.~Gu, C.~Zhu, C.~Ge, Y.~Zhang, X.~An, C.~Wang, Q.~Wang, X.~Liu, et~al.
\newblock Abdomenct-1k: Is abdominal organ segmentation a solved problem?
\newblock {\em IEEE Transactions on Pattern Analysis and Machine Intelligence}, 44(10):6695--6714, 2021.

\bibitem{milletari2016v}
F.~Milletari, N.~Navab, and S.-A. Ahmadi.
\newblock V-net: Fully convolutional neural networks for volumetric medical image segmentation.
\newblock In {\em 2016 fourth international conference on 3D vision (3DV)}, pages 565--571. Ieee, 2016.

\bibitem{UNet}
O.~Ronneberger, P.~Fischer, and T.~Brox.
\newblock U-net: Convolutional networks for biomedical image segmentation.
\newblock In {\em Medical Image Computing and Computer-Assisted Intervention--MICCAI 2015: 18th International Conference, Munich, Germany, October 5-9, 2015, Proceedings, Part III 18}, pages 234--241. Springer, 2015.

\bibitem{roy2023mednext}
S.~Roy, G.~Koehler, C.~Ulrich, M.~Baumgartner, J.~Petersen, F.~Isensee, P.~F. Jaeger, and K.~H. Maier-Hein.
\newblock Mednext: transformer-driven scaling of convnets for medical image segmentation.
\newblock In {\em International Conference on Medical Image Computing and Computer-Assisted Intervention}, pages 405--415. Springer, 2023.

\bibitem{setio2017validation}
A.~A.~A. Setio, A.~Traverso, T.~De~Bel, M.~S. Berens, C.~Van Den~Bogaard, P.~Cerello, H.~Chen, Q.~Dou, M.~E. Fantacci, B.~Geurts, et~al.
\newblock Validation, comparison, and combination of algorithms for automatic detection of pulmonary nodules in computed tomography images: the luna16 challenge.
\newblock {\em Medical image analysis}, 42:1--13, 2017.

\bibitem{UNETR++}
A.~Shaker, M.~Maaz, H.~Rasheed, S.~Khan, M.-H. Yang, and F.~S. Khan.
\newblock Unetr++: Delving into efficient and accurate 3d medical image segmentation.
\newblock {\em arXiv preprint arXiv:2212.04497}, 2022.

\bibitem{MSD}
A.~L. Simpson, M.~Antonelli, S.~Bakas, M.~Bilello, K.~Farahani, B.~Van~Ginneken, A.~Kopp-Schneider, B.~A. Landman, G.~Litjens, B.~Menze, et~al.
\newblock A large annotated medical image dataset for the development and evaluation of segmentation algorithms.
\newblock {\em arXiv preprint arXiv:1902.09063}, 2019.

\bibitem{taleb20203d}
A.~Taleb, W.~Loetzsch, N.~Danz, J.~Severin, T.~Gaertner, B.~Bergner, and C.~Lippert.
\newblock 3d self-supervised methods for medical imaging.
\newblock {\em Advances in neural information processing systems}, 33:18158--18172, 2020.

\bibitem{tang2022self}
Y.~Tang, D.~Yang, W.~Li, H.~R. Roth, B.~Landman, D.~Xu, V.~Nath, and A.~Hatamizadeh.
\newblock Self-supervised pre-training of swin transformers for 3d medical image analysis.
\newblock In {\em Proceedings of the IEEE/CVF Conference on Computer Vision and Pattern Recognition}, pages 20730--20740, 2022.

\bibitem{thawkar2023xraygpt}
O.~Thawkar, A.~Shaker, S.~S. Mullappilly, H.~Cholakkal, R.~M. Anwer, S.~Khan, J.~Laaksonen, and F.~S. Khan.
\newblock Xraygpt: Chest radiographs summarization using medical vision-language models.
\newblock {\em arXiv preprint arXiv:2306.07971}, 2023.

\bibitem{vaswani2017attention}
A.~Vaswani, N.~Shazeer, N.~Parmar, J.~Uszkoreit, L.~Jones, A.~N. Gomez, {\L}.~Kaiser, and I.~Polosukhin.
\newblock Attention is all you need.
\newblock {\em Advances in neural information processing systems}, 30, 2017.

\bibitem{TransBTS}
W.~Wang, C.~Chen, M.~Ding, H.~Yu, S.~Zha, and J.~Li.
\newblock Transbts: Multimodal brain tumor segmentation using transformer.
\newblock In {\em Medical Image Computing and Computer Assisted Intervention--MICCAI 2021: 24th International Conference, Strasbourg, France, September 27--October 1, 2021, Proceedings, Part I 24}, pages 109--119. Springer, 2021.

\bibitem{wang2023swinmm}
Y.~Wang, Z.~Li, J.~Mei, Z.~Wei, L.~Liu, C.~Wang, S.~Sang, A.~L. Yuille, C.~Xie, and Y.~Zhou.
\newblock Swinmm: masked multi-view with swin transformers for 3d medical image segmentation.
\newblock In {\em International Conference on Medical Image Computing and Computer-Assisted Intervention}, pages 486--496. Springer, 2023.

\bibitem{luo2022word}
L.~Xiangde, L.~Wenjun, X.~Jianghong, C.~Jieneng, S.~Tao, Z.~Xiaofan, et~al.
\newblock {WORD}: A large scale dataset, benchmark and clinical applicable study for abdominal organ segmentation from ct image.
\newblock {\em Medical Image Analysis}, 82:102642, 2022.

\bibitem{zhang2023dive}
C.~Zhang, H.~Zheng, and Y.~Gu.
\newblock Dive into the details of self-supervised learning for medical image analysis.
\newblock {\em Medical Image Analysis}, 89:102879, 2023.

\bibitem{nnFormer}
H.-Y. Zhou, J.~Guo, Y.~Zhang, L.~Yu, L.~Wang, and Y.~Yu.
\newblock nnformer: Interleaved transformer for volumetric segmentation.
\newblock {\em arXiv preprint arXiv:2109.03201}, 2021.

\bibitem{modelgenesis}
Z.~Zhou, V.~Sodha, J.~Pang, M.~B. Gotway, and J.~Liang.
\newblock Models genesis.
\newblock {\em Medical image analysis}, 67:101840, 2021.

\end{thebibliography}
\end{document}